\documentclass{elsart3}

\usepackage{graphicx}

\usepackage{amssymb}

\begin{document}

\begin{frontmatter}



\title{High resolution scanning tunneling spectroscopy \\
of ultrathin Pb on Si(111)-(6$\times$6) substrate}


\author{M. Krawiec},
\ead{krawiec@kft.umcs.lublin.pl}
\author{M. Ja{\l}ochowski \corauthref{cor1}}
\corauth[cor1]{Corresponding author, tel.: +48 81 537 6285,
               fax: +48 81 537 6191}
\ead{ifmkj@tytan.umcs.lublin.pl} and
\author{M. Kisiel}
\ead{mkisiel@tytan.umcs.lublin.pl}

\address{Institute of Physics and Nanotechnology Center, M. Curie-Sk\l odowska
University, pl. M. Curie Sk\l odowskiej 1, 20-031 Lublin, Poland}

\begin{abstract}
The electronic structure of Si(111)-(6$\times$6)Au surface covered with
submonolayer amount of Pb is investigated using scanning tunneling
spectroscopy. Already in small islands of Pb with thickness of $1$ ML
Pb$_{(111)}$ and with the diameter of only about $2$ nm we detected the
quantized electronic state with energy $0.55$ eV below the Fermi level.
Similarly, the I(V) characteristics made for the Si(111)-(6$\times$6)Au surface
reveal a localized energy state $0.3$ eV below the Fermi level. These energies
result from fitting of the theoretical curves to the experimental data. The
calculations are based on tight binding Hubbard model. The theoretical
calculations clearly show prominent modification of the I(V) curve due to
variation of electronic and topographic properties of the STM tip apex.
\end{abstract}

\end{frontmatter}

\section{Introduction}

Spectacular observations of electron confinement in Pb quantum wells (QW) have
been realized in several photoemission spectroscopy (UPS) experiments
\cite{Jal1}, \cite{Weitering}, \cite{Horn1}, helium atom scattering
\cite{Toennies}, \cite{Cvetko}, and in {\it in situ} electrical resistivity
measurements \cite{Jal2}. These techniques only provide surface electronic
structure information averaged over a large surface area. In contrast, the
scanning tunneling microscopy (STM), and scanning tunneling spectroscopy (STS)
allow one to study topographic and electronic structure with atomic scale
resolution \cite{Tromp}, \cite{Altfeder}, and \cite{Poelsema}. STS was also
successfully applied for studying the quantum size effects (QSE) in well
defined Pb islands, starting with the smallest thickness of 3 monoatomic layers
(ML) \cite{Tsong2} and laterally rather large. In our previous UPS study we
have shown that even 1 ML thick layer of Pb shows distinct QSE discrete
electronic level. However, clear experimental evidence directly relating
quantized electronic states with lateral size of individual islands has not
been reported. In the initial stage of ultrathin film growth, for very low
coverage, effects associated with formation of quantum dots are expected and
the STS technique is particularly well suited to detect and to study these
phenomena. However, quantitative analysis of tunneling current versus sample
bias I(V) dependence requires knowledge of the tunneling tip shape and its
electronic structure. Both parameters are in general unknown.

In this paper, we investigate the topographic and electronic structure of small
Pb islands and single Pb atoms on Si(111)-(6$\times$6)Au surface by means of
low temperature atomically resolved scanning tunneling spectroscopy. The
experimental studies are supplemented by theoretical ones based on a tight
binding model, where the I(V) characteristics are calculated using the
non-equilibrium Keldysh Green function formalism. We demonstrate how
(unavoidable during STS experiment) modification of the tunneling tip apex by
uncontrolled attaching or detaching of a single atom, modify the I(V) curve.
Moreover, we present existence of well defined QSE electronic level in 1 ML
thick Pb island with diameter smaller than 1.5 nm.

\section{Experimental method and results}
The experiment was carried out in UHV chamber equipped with an
Omicron variable temperature STM and RHEED. After preparation of
$Si(111)-(6\times6)Au$ surface under RHEED control at room
temperature (for further details of sample preparation see
\cite{Jal1}) the sample was transferred into a cooled STM stage
where a submonolayer amount of Pb was evaporated. STM tips were
produced by conventional electrochemical etching of tungsten wire,
and were further conditioned in {\it in situ} via prolonged scanning
over a clean Si(111)-(7$\times$7) surface with sample bias set
within the range from -5 to -10 V.  Tunneling spectroscopy was
performed simultaneously with topography measurements at every point
of the surface that is imaged by STM, or every fifth point sampled
during a scan. Typical STS data file contained 5000 I(V) curves with
200 I(V) points each.
The presented below STS curves are averages of several individual
curves collected within indicated areas. The number of these curves
spanned from 16 for smallest area, to 68 in the case of the largest
area. The area sizes were chosen in such a way that the individual
curves were very much the same shape, typically within $\pm$20 \% of
the tunneling current at any bias. This inaccuracy, due to
averaging, was further reduced. We stress that we have avoided
presenting of data from areas where shapes of neighboring I-V curves
were scattered and noisy. We are aware that averaging procedure
applied for sufficiently large number of curves (even if they differ
strongly) supplies nice, smooth curve.
The I(V) characteristics were acquired with feedback loop inactive.
During the measurements the temperature of the sample was about 130
K and the base pressure was less than 6$\times10^{-11}$\,mbar.

\begin{figure}[h]
\includegraphics[width=75mm]{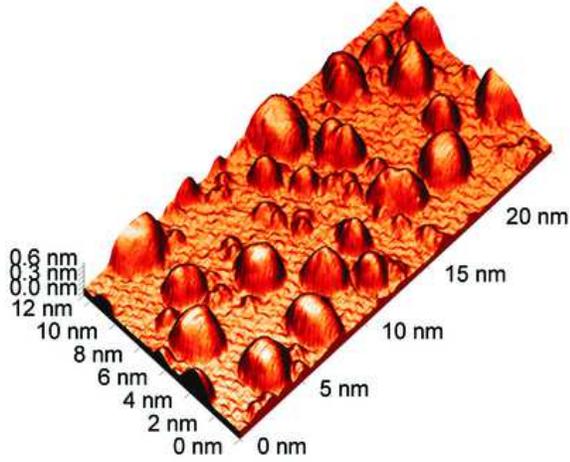}
 \caption{\label{Fig1} STM image of Si(111)-(6$\times$6)Au surface covered with
          0.2 ML of Pb. The sample bias voltage was -1.5 V and the tunneling
      current was 0.5 nA. The largest islands have thickness equivalent
      continuous Pb(111) ML. The smallest species visible are single Pb
      atoms. Periodic modulation of the Si(111)-(6$\times$6)Au
      reconstruction is also visible.}
\end{figure}

Figure \ref{Fig1} shows example of high resolution topographic data of the
Si(111)-(6$\times$6)Au surface with deposited 0.2 ML of Pb. The image shows
separate Pb islands with diameter ranging from about 1 nm to maximum of about
3 nm. At the sample bias equal to -1.5 V the height of the largest islands is
close to 0.4 nm. This corresponds to the height of a continuous monolayer of
Pb(111). The visible smallest species have the width and the height equal to 1
nm and 0.17 nm, respectively. The largest islands have flat tops whereas the
smaller appear as rounded features. We believe that the smallest features are
single Pb atoms, similarly as in the samples with coverage as low as 0.012 ML
of Pb(111) \cite{JalSurf}, whereas the largest one possesses crystalline
structure continuous monoatomic layer of Pb(111) \cite{Jal1}.

\begin{figure}[h]
\includegraphics[width=75mm]{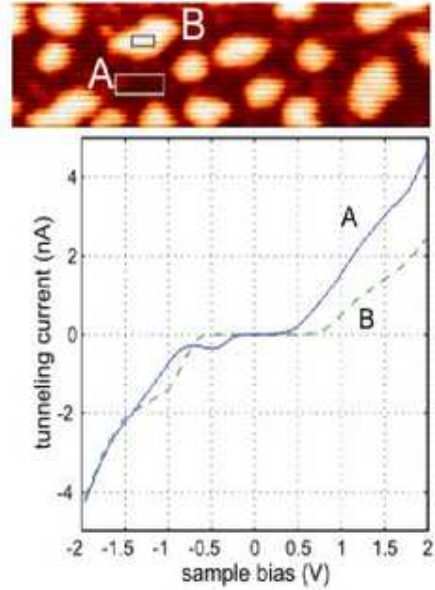}
\caption{\label{Fig2} 40nm$\times$10 nm (400$\times$100 pixels) STM image and
I(V) characteristics of Si(111)-(6$\times$6)Au surface covered with 0.2 ML of
the Pb. The characteristics were acquired every fifth pixel. The curves A and B
in the lower panel are averages over corresponding areas A and B shown in the
upper panel of the Figure. The feedback loop was opened at 3.1 nA and -1.76 V.}
\end{figure}

\begin{figure}[h]
\includegraphics[width=75mm]{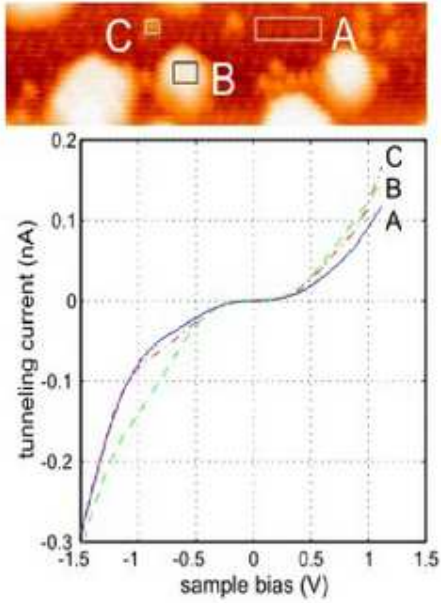}
\caption{\label{Fig3} 20nm$\times$5 nm (100$\times$25 pixels) STM image and I(V)
characteristics of Si(111)-(6$\times$6)Au surface covered with 0.2 ML of Pb. The
characteristics were acquired every pixel. The curves A, B, and C in the lower panel are
averages over corresponding areas A, B, and C shown in the upper panel of the Figure.
The feedback loop was opened at 0.5 nA and -1.76 V.}
\end{figure}

In comparison with the constant current imaging mode operation of
the STM, the spectroscopic mode requires cleanness and extra
stability of the tip. Although the STM tip was carefully conditioned
and its quality was checked during a control scan over freshly
prepared Si(111)-(7$\times$7) surface, we have seen frequently
reversible or irreversible changes of the tip properties. Although
these changes influenced only slightly the topographic images of the
surface, they have modified strongly the shape of the I-V tunneling
characteristics. This is shown in Fig. 2 and 3. In Fig. \ref{Fig2}
the tunneling characteristics taken on Si(111)-(6$\times$6)Au
surface exhibit negative differential resistance. We stress that
this phenomenon is related to the tip properties. In other
measurements, made with other tip, the region with negative
differential conductivity appeared only as shoulder shown in Fig.
\ref{Fig3}. In general the negative differential conductivity was
observed for the tips giving worse resolved topographic images.

%


\section{Theoretical description}
In general, the tunneling current depends on the Local Density of States (LDOS)
of both tip and the sample \cite{TersoffHamann}, but
in the most earlier experimental works no particular attention to the details
of the LDOS of the tungsten tip has been payed. In order to explain the
occurrence of the negative differential resistance and to correlate it with the
tip shape, we have developed a model of the tunneling system with the tip which
may attach a single atom or a cluster of atoms  - a case which occurs
frequently during scans.

\begin{figure}[h]
 \centering
 {
 \includegraphics[width=60mm]{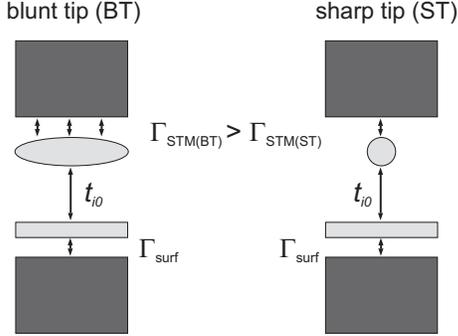}}
\caption{\label{Fig4} Schematic representation of two tips
considered in the tunneling current calculations.  Larger coupling
parameter $\Gamma_{STM}$ represents a large cluster of atoms with
bulk-like electronic structure. Smaller value of the coupling
parameter reflects localized, atomic-like character of a single atom
at the tip apex.}
\end{figure}

The system composed of surface, island and STM tip, Fig.\ref{Fig4},
is described by following Hamiltonian
\begin{eqnarray}
H = H_{STM} + H_{tip} + H_{isl} + H_{surf} + H_{int}, \label{Hamiltonian}
\end{eqnarray}
where
\begin{eqnarray}
H_{STM} = \sum_{{\bf k} \in STM} \epsilon_{\bf k} c^+_{\bf k} c_{\bf k} \label{H_STM}
\end{eqnarray}
and
\begin{eqnarray}
H_{surf} = \sum_{{\bf k} \in surf} \epsilon_{\bf k} c^+_{\bf k} c_{\bf k} \label{H_surf}
\end{eqnarray}
stand for the STM and the surface electrodes electrons with the energies
$\epsilon_{\bf k}$. The STM tip is modeled by a single atom with the energy
level $\varepsilon_0$
\begin{eqnarray}
H_{tip} = \varepsilon_0 c^+_0 c_0. \label{H_0}
\end{eqnarray}
Similarly the island is described by the energy level $\varepsilon_i$ %
\begin{eqnarray}
H_{isl} = \varepsilon_i c^+_i c_i. \label{H_i}
\end{eqnarray}
The interactions between different subsystems are in the form
\begin{eqnarray}
H_{int} = \sum_{{\bf k} \in STM} V_{{\bf k} 0} c^+_{\bf k} c_0 + t_{i0} c^+_0 c_i
\nonumber \\ + \sum_{{\bf k} \in surf} V_{{\bf k} i} c^+_{\bf k} c_i + H.c.
\label{H_int}
\end{eqnarray}
with $V_{{\bf k} 0}$ being a hybridization between the STM electrode and the
STM tip, $t_{i0}$ - the hopping integral between electrons on the island and
those in the tip, and $V_{{\bf k} i}$ - hybridization connecting the surface
and the island. Note that we have omitted the spin index in above equations as
in this case spin channels can be treated separately.

In order to calculate the STM tunneling current we follow the standard
procedure \cite{Haug} and the result reads
\begin{eqnarray}
I = \frac{2e}{\hbar} \int^{\infty}_{-\infty} \frac{dE}{2\pi} T(E) [f_{STM}(E) -
f_{surf}(E)], \label{curr}
\end{eqnarray}
where $f_{STM (surf)}(E)$ is the Fermi distribution function and the
transmittance $T(E)$ is given in the form
\begin{eqnarray}
T(E) = \Gamma_{STM}(E) \Gamma_{surf}(E) |G^r_{i 0}(E)|^2. \label{transmit}
\end{eqnarray}
$\Gamma_{STM}(E) = 2\pi \sum_{{\bf k} \in STM} |V_{{\bf k} 0}|^2 \delta{(E -
\epsilon_{\bf k})}$ and
$\Gamma_{surf}(E) =
2\pi \sum_{{\bf k} \in surf} |V_{{\bf k} i}|^2 \delta{(E - \epsilon_{\bf k})}$
is the coupling parameter between STM electrode and the tip atom and surface
and the island respectively. $G^r_{i 0}(E)$ is the the Fourier transform of the
retarded Green's function
$G^r_{i 0}(t) = i \theta(t) \langle [c_i(t), c^+(0)]_+ \rangle$ connecting the
tip atom $0$ with the island.

In numerical calculations we have chosen a constant density of
states in the STM electrode and the density of states in the surface
$\rho_{surf}(E) = |E+0.06|^{2.4}$. The parameter $t_{i0}$ is equal
to $2.5$ eV, which corresponds to the tip-surface distance $z
\approx 4$ \AA \cite{Calev,MK}. Such a small value of $z$ steams
from the fact that we have assumed the only single tunneling
channel. In realistic situation there are many of them and if we
take this effect into account the distance $z$ will be larger.

To make comparison to the experiment and reproduce the STM data shown in Figs.
\ref{Fig2} and \ref{Fig3}, we have used two different model tips. The tip is
described by a particle coupled to the STM electrode. Their coupling parameter
$\Gamma_{STM}$ depends on the tip
sharpness
and reflects a size of the particle attached to
the tip electrode.
The first one, which reproduces the
results shown in Fig. \ref{Fig2}, is characterized by a
particle with energy level
$\varepsilon_0 = 0.8$ eV strongly coupled to the STM electrode
$\Gamma_{STM} = 5$ eV. This corresponds to the real STM tip with small
curvature - blunt tip (BT), giving low resolution topographic images (see Fig.
\ref{Fig2}). The other one, reproducing the results shown in Fig. \ref{Fig3}
and giving high resolution topographic images, is a sharp tip (ST) modeled by
single atom with $\varepsilon_0 = 2.0$ eV weakly coupled to the STM electrode
$\Gamma_{STM} = 0.7$ eV.
Small value for the "sharp" tip describes localized, atomic-like
character of a single atom. Large coupling parameter represents a
large cluster of atoms with bulk-like electronic structure. This is
shown schematically in Fig.4.

%

\begin{figure}[h]
 {
 \includegraphics[width=70mm]{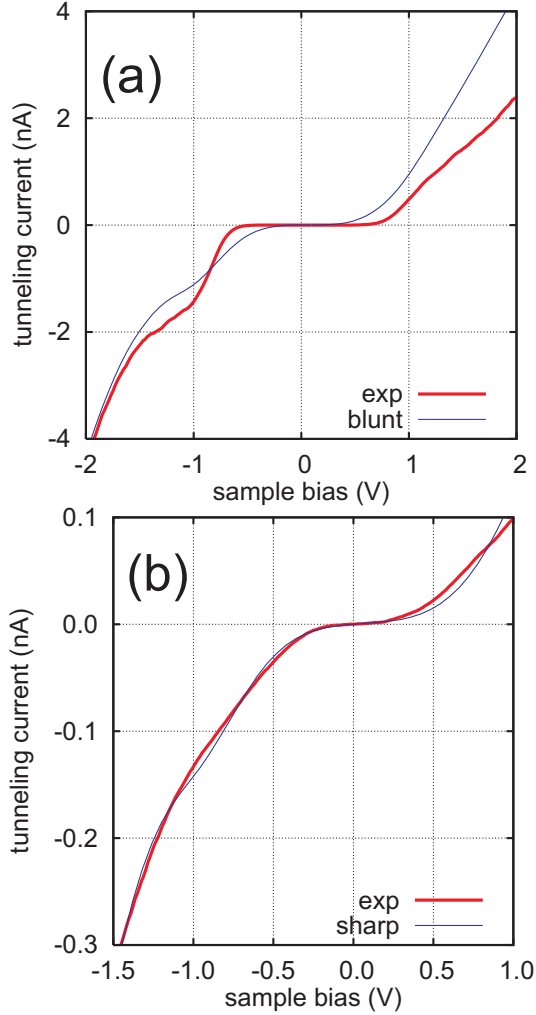}}
 \caption{\label{Fig5} The comparison of the I(V) experimental data of the
          tunneling current to the Pb island (thick line) with theoretical fit
      (thin line) for the blunt tip (a), and for the sharp tip (b),
      respectively. The model parameters are described in the text.}
\end{figure}

Figure \ref{Fig5}(a) shows the comparison of the experimental data
of the STM current-voltage characteristic of the Pb island (curve B
in Fig. \ref{Fig2}) and the theoretical calculations with a blunt
tip.
The theoretical fit has been done with assumption that the Pb island
is weakly coupled to the surface $\Gamma_{surf} = 0.6$ eV and a
single particle state $\varepsilon_i$ is at energy $-0.55$ eV with
respect to the Fermi energy. We identify this single particle state
energy with the quantum size state level in $1$ ML of Pb, as it was
determined in photoemission experiment \cite{Jal1}. The
corresponding comparison for a sharp tip (curve B in Fig.
\ref{Fig3}) is shown in Fig. \ref{Fig5}(b). Note that the parameters
characterizing the Pb island and the tip-surface distance are
exactly the same as those previously used. Such huge modifications
of the tunneling current are due to the properties of the tip only.

In order to reproduce the I(V) characteristics of the Si(111)-(6x6)Au surface,
we assumed that tunneling takes place into a small region of the surface,
containing one or a few atoms, artificially isolated from the rest of the
sample, which therefore can be also modeled as an island. This island is
characterized by energy level $\varepsilon_i = -0.3$ eV and the coupling to the
rest of the surface $\Gamma_{surf}$ is equal to $4$ eV. The single particle
energy level for Si(111)-(6x6)Au surface corresponds to energy of a flat
electron energy band found in photoemission experiment \cite{Hasegawa}. The
other parameters are the same as previously used.
The energies of the particles at the tip were chosen to fit the
calculated curves to the experimental ones. The only consequence of
their variation was change of the curves slope at higher biases but
not the positions of the I-V curves negative regions

\begin{figure}[h]
 {
  \includegraphics[width=70mm]{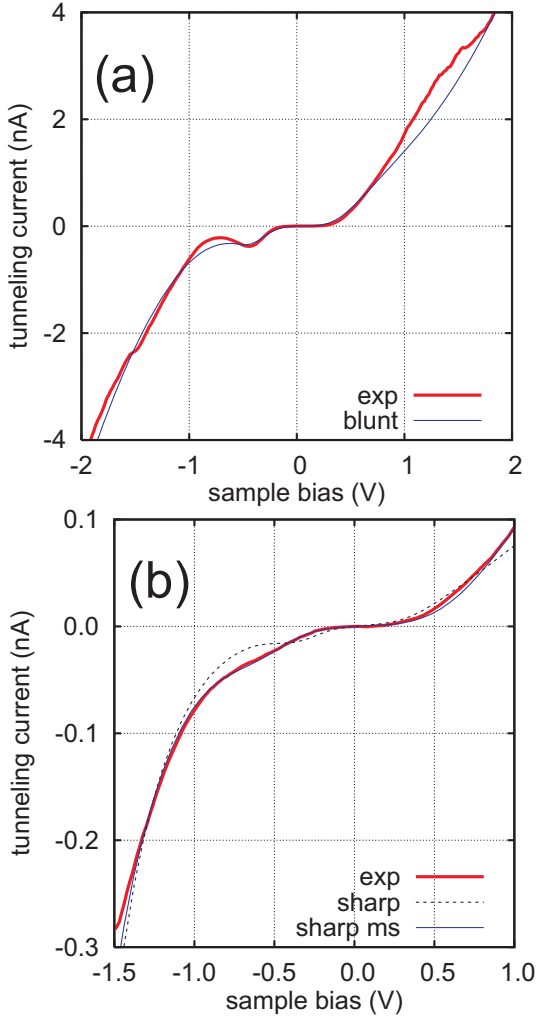}}
 \caption{\label{Fig6}The comparison of the I(V) data of the tunneling current
          to the Si(111)-(6$\times$6)Au island (thick line) with theoretical
      fit (thin line) for the blunt tip(a), and for the sharp tip (b),
      respectively. In (b) the dotted line corresponds to the coupling
      $\Gamma_{surf} = 4$ eV and the solid one is for $\Gamma_{surf} = 1$
      eV. The other parameters are described in the text.}
\end{figure}

The comparison of the theoretical calculations with the experimental
data is shown in Fig. \ref{Fig6}(a).
Note the negative behavior of the I(V) characteristic in a small
region below sample bias V $= -0.5$ V. Corresponding results for the
sharp tip are displayed in Fig. \ref{Fig6}(b).

If we use the same parameters characterizing the island as those in
Fig. \ref{Fig5}(b) and sharp tip, we get quite reasonable agreement
with the experimental data (dotted line), except for small region
around V $= -0.5$ V, which is the hallmark of the negative I(V)
characteristic observed in Fig. \ref{Fig6}(a).
\begin{figure}[h]
 \centering
 {
 \includegraphics[width=60mm]{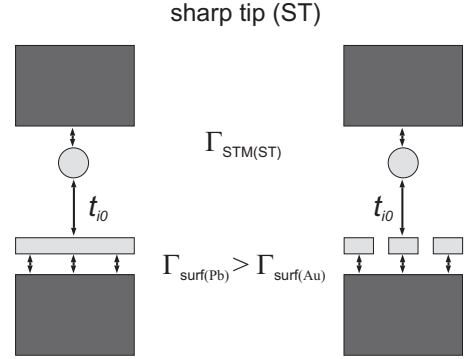}}
\caption{\label{Fig7} Schematic representation of two surfaces
modeled by islands. Larger coupling parameter $\Gamma_{surf(Pb)}$
represents Pb island. Smaller value of the coupling parameter
$\Gamma_{surf(Au)}$ describes small area of surface embedded in the
surrounding (6$\times$6)Au reconstruction and weakly coupled to the
substrate.}
\end{figure}
To improve this effect we had to make the coupling of the island to
the surface smaller. A four times smaller $\Gamma_{surf}$ gives
excellent agreement with the experimental results (thin solid line).
We believe that the use of the smaller value of $\Gamma_{surf}$ has
a physical origin and can be explained in the following way. For a
sharp tip the tunneling takes place to a narrower region of the
surface than in the case the blunt tip. This is well a understood
and accepted phenomenon. Therefore the surface region modelled by
the isolated island is also smaller, containing less Au atoms
(Fig.\ref{Fig7}). Further, if we assume that each atom on the island
is equally coupled to the surface we arrive at the conclusion that
in this case the coupling $\Gamma_{surf}$ should be smaller. Only
with this assumption we were able to get a perfect agreement with
the experiment. Note that in the case of the Pb island no such
modification of $\Gamma_{surf}$ is necessary, as the tunneling
region is bounded by the Pb island itself. As one can read off from
Fig. \ref{Fig1} the Pb islands are narrow and, more importantly,
quite high (a few of \AA), thus the tunneling directly into the
surface can be neglected.


\section{Conclusions}

In conclusion we have performed STS studies of ultrathin Pb on
Si(111)-(6$\times$6)Au surface supplemented by theoretical
calculations based on tight binding model. Already in small islands
of Pb with thickness of $1$ ML Pb$_{(111)}$ and with the diameter of
only about $2$ nm the quantized electronic state with energy $0.55$
eV below the Fermi level is detected. We identify this energy with
the quantum well state of 1 ML thick Pb island seen in the UPS
experiment \cite{Jal1}. Similarly, the I(V) characteristics made for
the Si(111)-(6$\times$6)Au surface reveal localized energy state
$0.3$ eV below the Fermi level, previously detected in photoemission
\cite{Hasegawa}. The obtained results lead also to the important
conclusion that measured I(V) characteristics should be taken with
care, as they strongly depend on the shape and the properties of the
tip itself, which is often omitted while discussing and interpreting
experimental data.

\noindent {\bf Acknowledgements} \\ This work has been supported by grant
no.1 P03B 004 28 of the Polish Committee of Scientific Research.

\end{document}